\documentclass[pra,floats,twocolumn,showpacs,superscriptaddress,floatfix]{revtex4-1}

\usepackage{graphicx}
\usepackage{ulem}
\usepackage{amssymb,amsmath,multirow,rotate,xcolor}
\usepackage{subfigure}
\usepackage{hyperref}

\newcommand{\alessio}[1]{{\color{blue} #1}}

\begin{document}

\title{Information sharing in Quantum Complex Networks}

\author{Alessio Cardillo}
\affiliation{Instituto de Biocomputaci\'on y F\'{\i}sica de Sistemas Complejos,
Universidad de Zaragoza, E-50018 Zaragoza, Spain.}
\affiliation{Departamento de F\'{\i}sica de la Materia Condensada, Universidad
de Zaragoza, E-50009 Zaragoza, Spain.}

\author{Fernando Galve}
\affiliation{Institute for Cross Disciplinary Physics and Complex Systems, 
IFISC (CSIC-UIB), Palma
de Mallorca, Spain}

\author{David Zueco}
\affiliation{Departamento de F\'{\i}sica de la Materia Condensada, Universidad
de Zaragoza, E-50009 Zaragoza, Spain.}
\affiliation{Instituto de Ciencia de Materiales de Arag\'on (ICMA),
  CSIC-Universidad de Zaragoza, E-50012 Zaragoza, Spain.}
\affiliation{Fundaci\' on ARAID, Paseo Mar\'{\i}a Agust\'{\i}n 36, E-50004
Zaragoza, Spain. }

\author{Jes\'us G\'omez-Garde\~nes}
\affiliation{Instituto de Biocomputaci\'on y F\'{\i}sica de Sistemas Complejos,
Universidad de Zaragoza, E-50018 Zaragoza, Spain.}
\affiliation{Departamento de F\'{\i}sica de la Materia Condensada, Universidad
de Zaragoza, E-50009 Zaragoza, Spain.}

\date{\today}

\begin{abstract}
We introduce the use of entanglement entropy as a tool for studying the amount of
information shared between the nodes of  quantum complex networks. By considering the ground state
of a network of coupled quantum harmonic oscillators, we compute the information
that each node has on the rest of the system. 
We show that the nodes storing the largest
amount of information are not the ones with the highest connectivity, but those
with intermediate connectivity thus breaking down the usual hierarchical picture
of classical networks.
We show both numerically and analytically that the mutual information
characterizes the network topology.
 As a byproduct, our results point out that the amount of
information available for an external node connecting to a quantum network
allows to determine the network topology.
\end{abstract}
\maketitle

\section{Introduction}

The advent of network science has influenced the research in many fields of
science in general, and physics in particular, in a pervasive way
\cite{barabasi}. Since the discovery of the structural features of real social,
biological and technological networks \cite{rev:albert,rev:newman}, the
development of the theoretical machinery of network science has blossomed  as an
efficient framework to interpret the many interaction patterns encoded in
real-scale complex systems of diverse nature \cite{newman_book} and to model
correctly the dynamical processes taking place on top of them
\cite{rev:bocc,vespignani}. 

One of the most important avenues of research in network science is its
connection with information theory. In this way, different
information-theoretical tools have been proposed to characterize the complexity
of networks beyond the typical statistical indicators such as their degree
distribution, clustering coefficient, degree correlations, etc
\cite{newman_book}. For instance, Shannon entropy, as shown in
\cite{GB1,GB2,GB3,Marro,Braunstein,GB0}, has been successfully applied to characterize the
complexity of ensembles of networks sharing some structural features while
information-theoretical tools have been also applied to the study of diffusion
processes on top of networks, such as random walks
\cite{ER1,ER2,ER3,GB,rosvall1,rosvall2}. 

The synergy between the field of complex networks and that of information theory has 
recently appealed to the quantum information community \cite{Ferraro, Garnerone2012b}. As a product,
classical results on percolation theory \cite{QPercAcin,QPercCuquet1,QPercCuquet2,QPercWu} and network
science, such as the small-world effect \cite{QSW}, have been revisited in networked structures of 
coupled quantum systems as a first step for designing quantum communication networks. 
Conversely, the use of quantum dynamical processes, such as quantum random walks \cite{muelken,Almeida2012} 
and their application to rank the importance of network elements \cite{QRsilvano,QRpaparo,QRburillo,Garnerone2012}, has given new quantum information perspectives 
to classical problems of the network realm.

The most fundamental characterization of a network is its connectivity distribution $P(k)$, i.e., the probability of finding a node connected to $k$ other nodes of the network. In addition to $P(k)$, many other statistical quantities in network scienceare used to characterize the topology relying on the sampling of the local measures (such as the degree $k$) of nodes \cite{newman_book}. On the other hand, quantum mechanical states built as ground states of many body Hamiltonians rely on both local and global lattice properties. As a consequence, the characterization of nodes' states in quantum complex network offers the possibility of extracting a novel characterization of nodes attributes, beyond those present in their local neighborhood.

In this work we  quantify the amount of mutual information that a single node shares with the rest of the network. To this aim, we compute the vacuum state of bosonic modes harmonically coupled through the specific adjacency matrix of a given complex network. We first show that the information contained in each node or lattice point is particularly characteristic (the precise meaning of which to be specified later) of the whole topology and, second\alessio{,} that hubs (nodes with the largest connectivity) become isolated in terms of the mutual information shared with the rest of the network. Both features are equally surprising from the
point of view of the classical network paradigm but, as we will discuss here, natural when quantum effects are incorporated. It is important to stress that both the models and the topologies studied here are far from solely being a fundamental curiosity. In fact, non regular topologies in quantum models, as those described here, are currently investigated in different contexts such as quantum emergent gravity models \cite{Caravelli2012}, Anderson localization \cite{Sade2005,Jahnke2008}, quantum phase transitions \cite{Halu2013} or optical communications \cite{Shapiro1982}.

\section{Theory and Model} As usual, we define a network as a set of $N$
nodes and $E$ edges (or links)  accounting for their pairwise interactions. The
network `backbone' is usually encoded in the Adjacency matrix, $A$, such that
$A_{ij}= 1$ if an edge connects nodes $i$ and $j$ while $A_{ij}=0$ otherwise. 
In this work we restrict to undirected networks so that $A_{ij} = A_{ji}$.
Although matrix $A$ stores all the structural meaning of a network it is more
convenient to rely on the so-called network Laplacian, $L$, to analyze its
structural and dynamical properties \cite{spectra}.  The Laplacian of a network
is defined from the Adjacency matrix as $L_{ij}= k_i \delta_{ij} - A_{ij}$,
where $k_i = \sum_j A_{ij}$ is the connectivity of node $i$, {\it i.e.}, the
number of nodes connected to $i$.  

A quite minimalistic manner to translate the features of a given Laplacian into a quantum system is to consider
identical, unit mass, quantum harmonic oscillators with equal on-site frequency
(normalized to $1$), and
interacting via springs as dictated by the adjacency matrix, the potential being $V=\sum_{i,j}cA_{i,j}(x_i-x_j)^2/2$. The resulting Hamiltonian
of the quantum network can be written then:
\begin{equation}
\label{Hnet}
H_{\rm network} =\frac{1}{2} \Big (   {\bf p}^{\rm T} {\bf p} + {\bf x}^{\rm T}
(\mathbb{I} + 2cL) {\bf  x}\Big )\, ,
\end{equation}
where $x_j = \frac{1}{\sqrt{2}}(a_j + a_j^\dagger)$ and $p_j = \frac{i}{\sqrt{2}}(a_j - a_j^\dagger)$
with the  bosonic anhilitation/creation operators satisfying the usual
commutation relations $[a_i, a_j^\dagger]=\delta_{ij}$ (we set $\hbar=1$).
As for a spring coupling matrix, the Laplacian guarantees $H_{\rm network} \geq 0$
and therefore the existence of a ground state.
Finally, we have included in the model a global coupling strength, $c$,
which is somehow arbitrary.  Our conclusions are independent of $c$ and it
can be seen as a regularization term \cite{fn1}.
%
The problem, despite quantum, is harmonic and therefore it is simple enough to
attack a complex topology and compute its ground state.  Almost any other 
Hamiltonian would make it impossible to perform the exact calculation of its ground
state in a complex topology.
Therefore Eq.~(\ref{Hnet}) is a minimal numerically solvable model. In
particular its ground state, the vacuum, relies on the 
eigenvectors of the Laplacian matrix and their corresponding eigenvalues shifted
by $1$. 
As a final note, let us comment that the classical limit of
(\ref{Hnet})  has a trivial ground state: all the nodes having $x_i=p_i= 0$
independent on the topology.
Therefore the quantum model, in this sense, is not trivial. Then all the ground state
mutual information, if any, shared by a node with the rest of the network is then
due to quantumness.

\begin{figure}[t!]
\centering
\includegraphics[width=0.85\columnwidth]{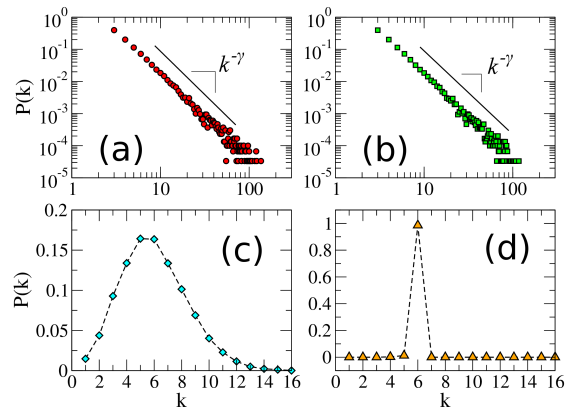}
\caption{(color online) Connectivity distribution $P(k)$ for scale
  free networks: Configurational (SF-CONF) (a) and Barab\'asi-Albert Scale-Free (SF-BA) (b).
  Figures (c) and (d) stand for Erd\H{o}s-R\'enyi (ER)   and Random Regular Graphs (RRG).}
\label{fig:1}
\end{figure}

We now quantify the amount of
information each of the elements of a network shares with the rest of the system. 
To this aim, we consider the partition of the network
into a node, say $i$, and its complement $i^{\rm c}$, {\it i.e.} the rest of the
network. Then, we compute the mutual information shared by the two parties as:
\begin{equation}
\label{mi}
{\mathcal{I}} (i | i^{\rm c} )= S_i + S_{i^{\rm c}} - S_{\rm tot}\;.
\end{equation}
Here $S_i$  and $S_{i^{\rm c}}$ are marginal (von Neumann) entropies and $S_{\rm tot}$ is the
total entropy of the network. 

Since the total network
is in its ground (hence pure) state we have $S_{\rm tot}=0$ and $S_i= S_{i^{\rm
c}}=\mathcal{I}(i|i^{\rm c})/2$. Therefore, the information that a node shares
with the network is intrinsically due to quantum correlations. Equivalently, the
mutual information is itself a measure of the entanglement (twice the entropy of entanglement, quantified by $S_i$) 
between a single node and the rest of the system. The marginal entropies for $i$ and $i^c$ read \cite{Agarwal1971}:
\begin{equation}
\!\!\!\!S_i=S_{i^{\rm
c}}=\Bigl(\!\mu_i+\frac{1}{2}\Bigr)\log\Bigl(\!\mu_i+\frac{1}{2}\Bigr)-\Bigl(\!\mu_i-\frac{1}{2}
\Bigr)\log\Bigl(\!\mu_i-\frac{1}{2}\Bigr)\,,
\label{entropy}
\end{equation}
which is a monotonically increasing function of $\mu_i$ that is characterized
by the second moments of the positions and momenta of nodes, $\mu_i= \sqrt
{\langle x_i^2 \rangle \langle p_i^2\rangle}$. After some algebra (see
Appendix \ref{appA}) we are able to quantify the value of $\mu_i$
as:
\begin{equation}\mu_i^2=\frac{1}{4}\sum_{j, j^\prime}S_{ij}^2 S_{ij^\prime}^2
\sqrt { \frac{1 + 2c\lambda_j}{1 + 2c\lambda_{j^\prime}} }\;,
\label{result}
\end{equation}
where $\{\lambda_j\}$ are the eigenvalues of the network Laplacian $L$ and
matrix $S$ accounts for the normal mode transformation that diagonalizes the
network Laplacian:  $L_d=S^{T} L S$ with  $S^T S = {\mathbb I}$.

From Eq.(\ref{entropy}) it is clear that each node has some mutual information
with the rest of the system provided $\mu_i>1/2$ whereas from Eq.(\ref{result})
we conclude that the amount of information depends on its contribution to each
of the Laplacian eigenvectors.  

In the following we quantify the entanglement entropies of nodes embedded in
different network topologies. First, we explore two homogeneous network
substrates: {\it (i)} Random Regular Graphs (RRG), in which all the nodes have
the same number of contacts ($k_i=\langle k\rangle, \forall i$), and {\it (ii)}
Erd\H{o}s-R\'enyi (ER) networks \cite{er}, for which the probability of finding a
node with $k$ neighbors, $P(k)$, follows a Poisson distribution so that most of
the nodes have a degree $k$ close to the average $\langle k\rangle$. Besides, we
have analyzed networks having a scale-free (SF) pattern for the probability
distribution, $P(k)\sim k^{-3}$, constructed by means of a configurational
random model (SF-CONF) \cite{conf}  and the Barab\'asi-Albert model (SF-BA)
\cite{ba} (see Figure \ref{fig:1}).



\section{Mutual information} 
We move to the numerical study and discussion of our results in all
those different complex network topologies. To this aim, we collect the
entanglement entropies of the $N_k$ nodes having connectivity $k$ and define the
average entanglement of the degree class $k$ as: $\langle
S_k\rangle=\sum_{i|k_i=k}S_i/N_k$.  One would expect that the larger the
connectivity $k$ of a node the more correlated it is with the rest of the network,
and thus the larger the value of $\langle S_k\rangle$. 

\begin{figure}
\includegraphics[width=1.00\columnwidth]{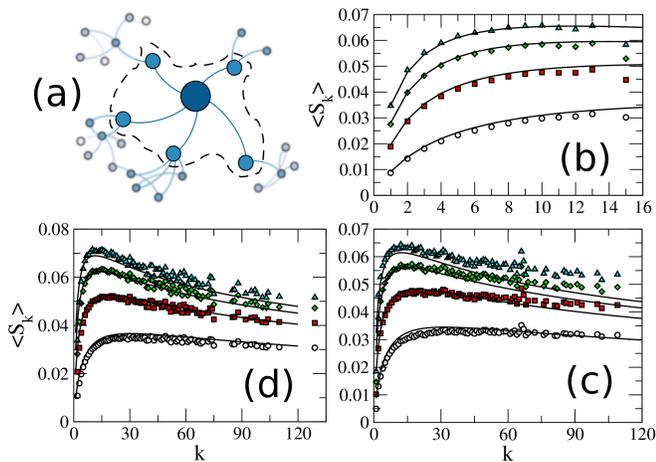}
\caption{(color online) In {\bf (a)} we show the microscopic picture: a
particular node and its boundary (first neighbors) with the rest of the system
(blurred). The rest of the panels show the average entropy of nodes with degree
$k$, $\langle S_k \rangle$ for the following topologies: {\bf (b)} ER, {\bf (c)}
SF-CONF and {\bf (d)} SF-BA networks. Each symbol refers to a value of the
coupling strength $c$. In particular, we have that
$\bullet$,$\blacksquare$,$\blacklozenge$,$\blacktriangle$ correspond to $c=
0.2$, $0.4$, $0.6$ and $0.8$ respectively. All the networks have the same
average degree $\langle k \rangle = 4$. Solid lines represent the theoretical
curves calculated using the mean-field formulation.
The fitted values for $\kappa$ are $\kappa =3.8, 3.5, 4.2$ for the ER,
SF-BA and SF-CONF respectively. All the results are averaged over 30 realizations of each kind of network.
}
\label{fig:2}
\end{figure}

\begin{figure}
\centering
\includegraphics[width=1.0\columnwidth]{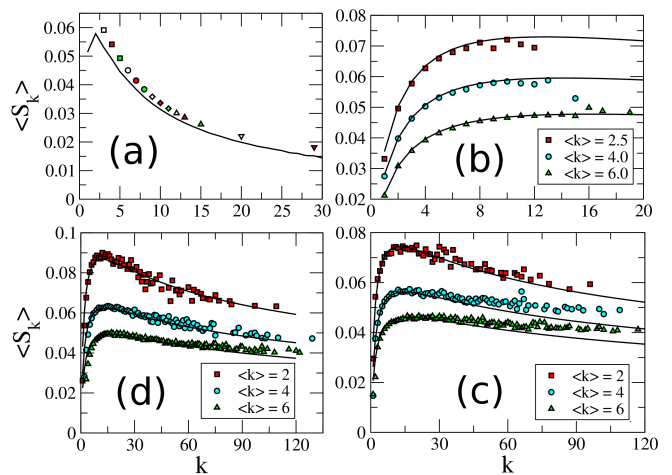}
\caption{(color online) Average entropy of nodes with degree $k$, $\langle S_k
\rangle$, for all the topologies under study: {\bf (a)} RRG , {\bf (b)} ER, 
{\bf (c)} SF-CONF and {\bf (d)} SF-BA networks. All the oscillators are coupled
with the same strength $c= 0.6$. Each symbol represents a different value
of the average degree $\langle k \rangle$ of the system. Solid lines represent the
theoretical behavior calculated with the mean-field formulation.
The fitted values for$\kappa$ for ER $\kappa = 2.6, 3.8, 5.5$, for
SF-BA $\kappa =1.8, 3.5, 5.2$ and SF-CONF $\kappa =2.5, 4.2, 5.8$. 
All the results are averaged over 30 realizations of each kind of network.
}
\label{fig:3}
\end{figure}

The panels in Figs.~\ref{fig:2} and \ref{fig:3} summarize  our findings for the
behavior of $\langle S_k\rangle$.  In Fig,~\ref{fig:2} we explore ER [panel {\bf
(b)}], SF-CONF [panel {\bf (c)}] and SF-BA [panel {\bf (d)}] networks for
different values of the coupling $c=0.2,0.4,0.6$ and $0.8$. For ER networks we
observe that the value of $\langle S_k\rangle$ increases with $k$ although we
note that the growing trend seems to saturate for large values of $k$ pointing
out that entanglement is bounded. On the other hand for SF networks, displaying
a larger heterogeneity for the collection of degrees, the growing trend of
$\langle S_k\rangle$ only holds for small to moderate connectivities $k$, then
$\langle S_k\rangle$ reaches a maximum and starts to decrease. Eventually, those
nodes with sufficiently large $k$ would drop its entanglement entropy. As a
result {\it  hubs} are not the most entangled nodes, but there is an
optimally-correlated class of nodes having moderate connectivity. The plots in
Fig.~\ref{fig:3} confirm the above results. In these cases, we have fixed $c=0.6$ and
changed the mean connectivity of the RRG [panel {\bf (a)}], ER [panel {\bf
(b)}], SF-CONF [panel {\bf (c)}]  and SF-BA [panel {\bf (d)}] networks.
 Notice that in the RRG the mean connectivity and the connectivity are the same (all 
 the nodes have the same connectivity, see Fig. \ref{fig:1}). Therefore, considering a net with 
 some $\langle k\rangle$ would provide only one point in the curve $\langle S_k\rangle$,
 so we merge in one plot different networks with different connectivities.

\section{Mean-field formulation} In the following we develop a minimal model aimed
at capturing the rise-and-fall behavior of $\langle S_k\rangle$ in SF networks. 
The simplest framework to deal with bipartite entanglement is
sketched in  Fig.~\ref{fig:2}.{\bf (a)}. Rather than selecting a site $i$ and replacing the interaction with each of its neighbors by a
mean value (standard mean field), we consider both $i$ and its neighbors. The mean field approximation enters when replacing the interaction of the neighbors of $i$ with their corresponding neighbors by its mean value. This mean field assumption is equivalent to a renormalization of the frequency of the $k$ neighbors of $i$. In this way, the mean field Hamiltonian for a node $i$ with $k$ neighbors yields:
\begin{equation}
H_{\rm MF}^k=\frac{1}{2} ( p_i^2+x_i^2 )
+\frac{1}{2}\sum_{j=1}^k p_j^2 + \nu^2_{\kappa}x_j^2 + c(x_i-x_j)^2
\, ,
\label{Hmf}
\end{equation}
where the renormalized frequency of the neighbors reads:
$\nu^2_{\kappa}=1+2c\kappa$, and $\kappa$ is a fitting parameter (see below).

This model is analytically solvable (see Appendix \ref{appB}).
Therefore, we can find the entropy of the central node $\langle S_k\rangle^{MF}$
analytically as a function of its connectivity $k$ and $\kappa$. In
Figs.~\ref{fig:2} and \ref{fig:3} we plot with solid lines the curves $\langle
S_k\rangle^{MF}$ obtained  after tuning the single parameter $\kappa$ for each
of the curves. Our mean-field approximation agrees
fairly good with the numerics. 
The values of $\kappa$ for the different topologies under study are shown in Table~\ref{tab:kappa}. As observed, in the case of Erd\H{o}s R\'enyi (ER) networks with average degree $\langle k \rangle = 2$ results are missing. This is due to the fact that for average degrees lesser than 2.5 the resulting ER networks are not made by a unique connected component. Importantly, we observe that $\kappa$ only depends on the average degree and on the considered underlying topology as it is independent of the coupling
parameter $c$ (see Figs. \ref{fig:2} and \ref{fig:3}). 

In addition to the quantitative agreement, the analytical estimation $\langle
S_k\rangle^{MF}$ allows to explain the rise-and-fall of entropy across degree
classes.
%
%
\begin{table}[hb!]
\setlength{\tabcolsep}{6.5pt}
\centering
\begin{tabular}[c]{|l|c|c|c|c|}\hline
\textbf{$\langle k \rangle$ \textbackslash  $c$} & \textbf{RRG} & \textbf{ER} & \textbf{SF-BA} & \textbf{SF-CONF}\\
\hline
2 \textbackslash 0.2 & -- & 2.6 & 1.8 & 2.5\\
\hline
2 \textbackslash 0.4 & -- & 2.6 & 1.8 & 2.5\\
\hline
2 \textbackslash 0.6 & -- & 2.6 & 1.8 & 2.5\\
\hline
2 \textbackslash 0.8 & -- & 2.6 & 1.8 & 2.5\\
\hline
4 \textbackslash 0.2 & 3.4 & 3.8 & 3.5 & 4.2\\
\hline
4 \textbackslash 0.4 & 3.4 & 3.8 & 3.5 & 4.2\\
\hline
4 \textbackslash 0.6 & 3.4 & 3.8 & 3.5 & 4.2\\
\hline
4 \textbackslash 0.8 & 3.4 & 3.8 & 3.5 & 4.2\\
\hline
6 \textbackslash 0.2 & 5.2 & 5.5 & 5.2 & 5.8\\
\hline
6 \textbackslash 0.4 & 5.2 & 5.5 & 5.2 & 5.8\\
\hline
6 \textbackslash 0.6 & 5.2 & 5.5 & 5.2 & 5.8\\
\hline
6 \textbackslash 0.8 & 5.2 & 5.5 & 5.2 & 5.8\\
\hline
\end{tabular}
\caption{Fitting parameter $\kappa$ for all the different network
  topologies with respect to each pair of parameters, namely: average degree $\langle k \rangle$ and coupling $c$.}
\label{tab:kappa}
\end{table}
As shown in the Appendix \ref{appB}, this phenomenon lies
in the fact that hubs are {\it almost} eigenvectors of the Laplacian and thus
normal modes of the Hamiltonian (uncoupled from the rest of the system and therefore not entangled).
The progressive localization of the eigenvectors with $k$ balances the growth of the correlations associated to the increase of $k$. It is the competition between these two effects what explains
the peak for $\langle S_k\rangle$ in SF networks at moderately, rather that
maximally, coupled nodes.
\begin{figure}
\centering
\includegraphics[width=0.850\columnwidth]{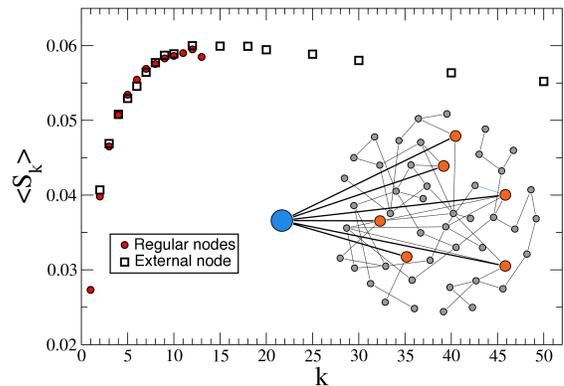}
\caption{(color online) Entropy share by an external (blue) node coupled to an ER
graph. In the plot  we show (filled dots) the entanglement entropy $\langle
S_k\rangle$ of the nodes of the ER network as a function of their connectivity.
In addition, we show (empty squares) the evolution of the entanglement of the
external node as a function of its connectivity, {\it i.e.} the number of
links launched to the target network.}
\label{fig:4}
\end{figure}

\section{Attaching an external node}
Let us now tackle the problem from a different perspective. Up to this point we have assumed that we have access to any of the nodes and thus computed their corresponding mutual information to characterize the network. Now, we consider the network as an unknown system and aim at recovering the above results by using a single node that can be attached to the network with as many links as desired. If such node/probe could get further information about its own entropy, say by measuring its purity, it could sequentially connect to more and more nodes in a random way so to reproduce a  curve $\langle S_k\rangle$. 
We have represented this situation in Fig.~\ref{fig:4}, for an ER graph, as compared to the situation in which no probe is present. As shown, the external probe realizes, by launching more than $k\simeq 15$ links,
that the amount of information it can extract from a the network is bounded and its maximum is reached by means of a moderate number of connections. We note that this individual entropy fairly coincides with the entropy that any internal node of the same connectivity would measure. 

\begin{figure}[!t]
\centering
\includegraphics[width=1\columnwidth]{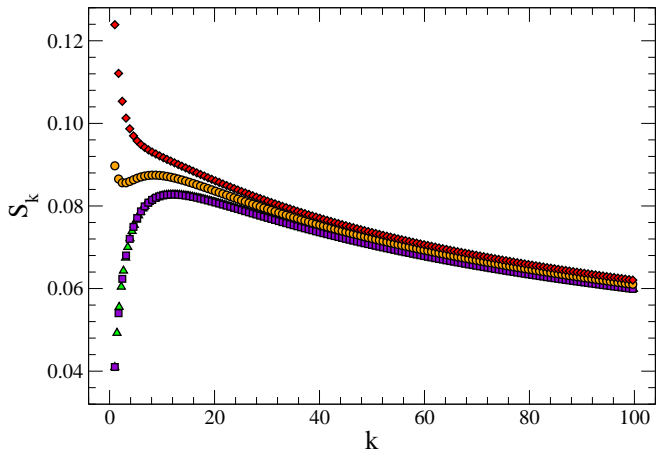}
\caption{(color online) Rise and fall behavior for finite temperature. We plot the expected $S_k$
for our mean field description with the values $c=0.6$ and $\kappa=2$, 
and temperatures $T =0.001$ (green $\blacktriangle$), $0.1$ (purple $\blacksquare$), $0.25$ (orange $\bullet$), $0.3$ (red $\blacklozenge$).}
\label{fig:5}
\end{figure}

\section{Robustness against temperature}
We use our mean field approach to explore the effect of temperature on the observed behavior at zero temperature. 
Instead of assuming the ground state of the network, we take a Gibbs
density matrix $\rho=\exp(-H_{\rm net}/T)/Z$ with temperature $T$, 
 $Z= {\rm Tr}[\exp(-H_{\rm net}/T)]$ the partition function and we
 have set $k_{\rm B}=1$.
Figure~\ref{fig:5} shows the result for $c=0.6$ and fitting parameter $\kappa=2$: the rise and fall behavior survives for $T=0.1$  and starts
to disappear only when $T>0.25$. Furthermore, nodes with
high connectivity are left almost unaffected by a temperature increase. Indeed, highly connected nodes become almost eigenmodes of
the system and their frequency gets renormalized by $k$, meaning that their frequency is very high as compared to temperature. Effectively
they do not feel the temperature increase. In contrast, nodes with low connectivity do not feel this renormalization so much, thus leading
to the typical increase of entropy due to temperature. In conclusion, the fact that for $T =0.1$ the curve matches
the zero temperature one confirm the desired stability for our results and conclusions.

\section{Conclusions}
The entropy of entanglement of nodes in quantum oscillators networks reveals a novel and non trivial
  characterization of single node attributes. In particular, the decay
  of the entanglement for large connectivity nodes is seen as the
  fingerprint of the localization of some of the Laplacian
  eigenvectors around hubs which turns them into normal modes of the
  system. This effect balances the increase of the entanglement with
  the connectivity, analogously to an area law in regular lattices,
  thus causing the rise-and-fall of the entanglement entropy across
  connectivity classes. 
We further stress that the results presented here survive in presence of a finite, but small
  temperature.

 In addition to their interest for the emerging field of quantum information on networks, our results show an interesting connection with fundamental concepts of quantum gravity in complex spatial connectivities. In fact, the setup used here can be seen as the discretized version of real massive Klein-Gordon fields far from the usual  Minkowsky or curved space-time situations, suggested from
emergent gravity concepts as intermediate topologies in the transition from a
highly connected (high energy) quantum geometric phase of the universe to the
low energy, largely homogeneous, actual phase \cite{smolin}. If the links of the network (related to
the quantum gravitational field) are seen as a heat bath for quantum fields `living' on it, the effect of 
different entropy densities could lead to entropic forces, and therefore to preference of some topological configurations over others. Although the connection between complex space-time topologies and the field of network science has recently attracted attention \cite{cosmo}, any result coming from this synergy
has to be considered preliminary and thought-provoking.

\acknowledgments	
We acknowledge Diego Blas for discussions and support from the Spanish DGICYT under projects, FIS2011-14539-E (EXPLORA program), FIS2011-23526 and FIS2011-25167, and by the Arag\'on (Grupo FENOL) and Balearic Goverments. F.G. acknowledges the CSIC post-doctoral JAE program and TIQS (FIS2011-23526) project. J.G.G. is supported by MINECO through the Ram\'on y Cajal program.

\appendix

\section{Entanglement entropy}
\label{appA}

Let us start by deriving Eq. (4) in the text. Given $\varrho$,  the quantum state of the network,  its associated Von Neumann entropy is given by,
\begin{equation}
\label{VN}
S = -{\rm Tr} ( \varrho \ln \varrho )\;,
\end{equation}
where ${\rm Tr}$ accounts for the trace operation. The marginal entropy for the node $i$, $S_i$, is obtained by replacing $\varrho$ in (\ref{VN}) by the reduced density matrix, 
\begin{equation}
\varrho_i = {\rm Tr}_{i^{\rm c}} \varrho \; ,
\end{equation}
where ${\rm Tr}_{i^{\rm c}}$ is the partial trace, {\it i.e.}, the trace over the complement of $i$ (the rest of the nodes).

In our work, the state of the network considered is the ground state of the Hamiltonian [Eq. (1) in the text]:
\begin{equation}
\label{Hnet_app}
H_{\rm network} =\frac{1}{2} \Big (   {\bf p}^{\rm T} {\bf p} + {\bf x}^{\rm T}
(\mathbb{I} + 2cL) {\bf  x}\Big )\, ,
\end{equation}
here $\mathbb{I}$ is the $N \times N$ identity matrix, $c$ is the coupling strength between connected oscillators and $L$ the network Laplacian.  The operators  ${\bf p}^{\rm T}= (p_1,  p_2,..., p_N)$ and ${\bf x}^{\rm T}= (x_1, , x_2,...,x_N)$ are the momenta and positions of nodes respectively, satisfying the usual commutation relations:
$[{\bf x}, {\bf p}^{\rm T}] = {\mbox i}\hbar\,\mathbb{I}$.
We are interested in analyzing the  ground state of the system, which is a pure state, thus having  $S =
0$. On top of that, this state is Gaussian (since the Hamiltonian is quadratic) so that the reduced density matrices $\varrho_i$ and $S_i$ can be computed by means of the covariance matrix:
\begin{equation}
\sigma=\left (
\begin{array}{cc}
\langle x_i^2 \rangle & \frac{1}{2} \langle x_i p_i + p_i x_i \rangle
\\
\frac{1}{2} \langle x_i p_i + p_i x_i \rangle & \langle p_i^2 \rangle
\end{array}
\right )\;,
\end{equation}
where the averages are calculated via the reduced density matrix, $\varrho_i$, as: $\langle x_i^2
\rangle = {\rm Tr} (x_i^2 \varrho_i)$. It was Agarwal \cite{Agarwal1971} who derived an explicit formula for the marginal entropies [Eq. (3) in the text]:
\begin{equation}
\!\!\!\!S_i=S_{i^{\rm
c}}=\Bigl(\!\mu_i+\frac{1}{2}\Bigr)\log\Bigl(\!\mu_i+\frac{1}{2}\Bigr)-\Bigl(\!\mu_i-\frac{1}{2}
\Bigr)\log\Bigl(\!\mu_i-\frac{1}{2}\Bigr)\,,
\label{entropy-App}
\end{equation}
with $\mu_i= \sqrt
{\langle x_i^2 \rangle \langle p_i^2\rangle}$.

We are able to find these quadratures by working with normal modes, {\it i.e.}, those diagonalizing
the potential energy matrix $V=\mathbb{I}+2cL$:
\begin{equation}
{\bf x} = S \, {\bf Q}\ \textrm{so that} \ S^{\rm T} V S =V_d,
\end{equation}
whose quadratures are those of a set of uncoupled oscillators at their individual ground state: 
\begin{eqnarray}
\langle Q_i\rangle =\langle P_i\rangle&=&0 
\\
 \langle Q_iQ_j\rangle&=&\delta_{ij}\frac{\hbar}{2 \Omega_j}
\\
 \langle
 P_iP_j\rangle&=&\delta_{ij}\frac{\hbar\Omega_j}{2 }
\\
\frac{1}{2}\langle Q_i P_j +Q_jP_i\rangle&=&0,
\end{eqnarray}
with $\Omega_i=\sqrt{1+2 c \lambda_i}$ the eigenfrequencies, and $\lambda_i$ the eigenvalues of the
Laplacian matrix $L$. The latter relation is obtained from simple inspection of the eigenvalue equation $V{\bf u}=\Omega^2{\bf u}=(\mathbb{I}+2cL){\bf u}$, so that $L{\bf u}=\frac{\Omega^2-1}{2c}{\bf u}\equiv \lambda{\bf u}$. That is, ${\bf u}$ is an eigenvector of both $V$ and $L$, with $\Omega=\sqrt{1+2c\lambda}$. Then we obtain:
\begin{align}
\label{qi}
\langle x_i \rangle
&=
\sum_{j} S_{ij} \langle Q_j \rangle = 0
\\
\label{pi}
\langle p_i \rangle
&=
\sum_{j} S_{ij} \langle P_j \rangle = 0
\\
\label{qipi}
\langle p_i x_j \rangle
&=
\sum_{k,l} S_{ik}S_{jl}  \langle P_k Q_l \rangle = 0
\\
\label{qi2}
\langle x_i^2 \rangle
&=
\sum_{j} \big ( S_{ij} \big )^2 \langle Q_j^2 \rangle
=
\sum_{j} \big ( S_{ij} \big )^2 \frac{\hbar}{2 \Omega_j}
\\
\label{pi2}
\langle p_i^2 \rangle
&=
\sum_{j} \big ( S_{ij} \big )^2 \langle P_j^2 \rangle
=
\sum_{j} \big ( S_{ij} \big )^2 \frac{\hbar \Omega_j}{2}.
\end{align}
Finally, we arrive to the expression for the quadratures:
\begin{eqnarray}
\mu_i^2&=&\langle x_i^2\rangle\langle p_i^2\rangle-\frac{1}{2} \langle x_i p_i + p_i x_i \rangle\nonumber\\
&=&\langle x_i^2\rangle\langle p_i^2\rangle=\frac{1}{4}\sum_{j} \big ( S_{ij} \big )^2\big ( S_{ij'} \big )^2 \frac{\Omega_j}{\Omega_{j'}},
\end{eqnarray}
as stated in the main text.

\section{Entanglement mean field approximation}
\label{appB}
In this section we sketch the solution for the mean field
approximation presented in the main text.
The mean-field Hamiltonian for a node surrounded by $k$ neighbors
[Cf. Fig. \ref{fig:1}{\bf a}] can be rewritten in matrix form like as:
\begin{equation}
\label{Hmf}
H=\frac{1}{2}\Big ({\bf p}^{\rm T} \, {\mathbb I}\, {\bf p}+{\bf x}^{\rm T} \, \hat V \, {\bf x}\Big )
\end{equation}
whith the (k+1)-tuples:
\begin{equation}
\bf p=\left (\begin{array}{c} p_0
\\
p_1
\\
\vdots
\\
p_k
\end{array}
\right )\; ,\qquad\bf x =\left (
\begin{array}{c}
x_0
\\
x_1
\\
\vdots
\\
x_k
\end{array}
\right ) \, ,
\end{equation}
note that we have named the 0-node the central one.  The then potential reads:
\begin{equation}
\label{V}
\hat V
=
\left (
\begin{array}{ccccc}
1 + c \, k & -c & -c &  \cdots &-c
 \\
 -c & 1 + c \, \kappa & 0& \cdots & 0
 \\ 
 -c &  0  & 1 + c \, \kappa   & \cdots & 0
 \\
 \vdots & \vdots &   & \ddots & \vdots
 \\
 -c & \cdots & \cdots & 0 & 1 + c \, \kappa
\end{array}
\right ).
\end{equation}

The equilibrium properties of (\ref{Hmf}), in particular the Von Neumann
entropy, is characterized by the eigenvalues and eigenvectors of
$\hat V$, as explained in the previous section.  It turns out that the spectrum of $\hat V$ given by
(\ref{V}) can be analytically computed:
\begin{itemize}
\item [\it i)]
The $(k+1)\times (k+1)$ matrix $\hat V$  in (\ref{V}) has $(k+1)-2$
eigenvectors of the form:
\begin{equation}
| \lambda_j \rangle 
=
\left (
\begin{array}{c}
0
\\
\vdots
\\
1_i
\\
\vdots
\\
-1_j
\\
\vdots
\end{array}
\right ),
\end{equation}
with degenerated eigenvalues
\begin{equation}
\lambda_j = 1+ c \kappa
\end{equation}
as can be easily checked.
\item [{\it ii)}]
The other two eigenvectors are of the form,
\begin{equation}
| \lambda\rangle 
=
\frac{1}{\sqrt{z^2 + k}}
\left (
\begin{array}{c}
z
\\
1
\\
\vdots
\\
1
\end{array}
\right )
\end{equation}
where the eigenvalues and eigenvectors are found from the equations,
\begin{align}
(\omega + c k^2) - k c &= \lambda z\;,
\\
-c z + \omega+  c \kappa &= \lambda\;,
\end{align}
with eigenvalues
\begin{equation}
\lambda_\pm
=
\frac{1}{2} \left\{c \left[\kappa \pm\sqrt{\kappa ^2-2 \kappa k+k (k+4)}+k\right]+2\omega \right\}
\end{equation}
and
\begin{equation}
z_\pm= \frac{1}{2} \left[\kappa - k \mp \sqrt{\kappa ^2-2 \kappa k+k (k+4)}\right].
\end{equation}
\end{itemize}  
The latter are the only ones entering in the formula for the
marginal entropy of the node 0, see (\ref{entropy-App}) and Eq. (3) in
main text.  Thus, the quadratures can be written as:
\begin{align}
\langle x_0^2 \rangle
&= 
\sum_{i = \pm}
\frac{z_i^2}{z_i^2 + k}
\frac{1}{2 \sqrt {\lambda_i}}\;,
\\
\langle p_0^2 \rangle
&= 
\sum_{i = \pm}
\frac{z_i^2}{z_i^2 + k}
\frac{\sqrt {\lambda_i}}{2}\;,
\end{align}
from which the entropy is obtained.

We finally note that $z_+ \to -k$ for large enough $k$.
Therefore in this limit the corresponding eigenvector approaches to
$(1, 0, ..., 0)$ with
frequency $\omega = \sqrt{\lambda_+} \to \sqrt{1 + 2 c k}$.
Therefore, in the limit of large connectivity the node is a normal mode
and its corresponding marginal entropy approaches to zero.

\medskip
\nocite{*}

\bibliographystyle{apsrev4-1}

\begin{thebibliography}{99}

\bibitem{barabasi} A.L. Barab\'asi, Nature Phys. {\bf 8}, 14 (2012).

\bibitem{rev:albert} R. Albert and A.-L. Barab\'asi,  Rev. Mod. Phys. {\bf 74},
47 (2002).

\bibitem{rev:newman} M.E.J. Newman, SIAM Rev. {\bf 45}, 167 (2003).

\bibitem{newman_book} M.E.J. Newman, {\it Networks: An Introduction}, Oxford University Press, USA (2010).

\bibitem{rev:bocc} S. Boccaletti, V. Latora, Y. Moreno, M. Chavez and D.-U. Hwang,  Phys. Rep. {\bf 424}, 175 (2006).

\bibitem{vespignani} A. Vespignani, Nature Phys. {\bf 8}, 32 (2012).


\bibitem{GB1} G. Bianconi, EPL {\bf 81}, 28005 (2008).

\bibitem{GB2} G. Bianconi, Phys. Rev. E {\bf 79}, 036114 (2009).

\bibitem{GB3} K. Anand and G. Bianconi, Phys. Rev. E {\bf 80} 045102 (2009).

\bibitem{Marro} S. Johnson, J.J. Torres, J. Marro and M.A. Mu\~noz, Phys. Rev. Lett. {\bf 104}, 108702 (2010).

\bibitem{Braunstein} S.L. Braunstein, S. Ghosh, T. Mansour, S. Severini and R.C. Wilson, Phys. Rev. A {\bf 73}, 012320 (2006).

\bibitem{GB0} K. Anand, G. Bianconi and S. Severini, Phys. Rev. E {\bf 83}, 036109 (2011).	



\bibitem{GB}  G. Bianconi, P. Pin and M. Marsili, Proc. Nat. Acad. Sci. (USA) {\bf 106}, 11433 (2009).


\bibitem{ER1} J. G\'omez-Garde\~nes and V. Latora, Phys. Rev. E {\bf 78}, 065102 (2008). 

\bibitem{ER2} Z. Burda, J. Duda, J.M. Luck and B. Waclaw, Phys. Rev. Lett. {\bf 102}, 160602 (2009).

\bibitem{ER3} R. Sinatra, J. G\'omez-Garde\~nes, R. Lambiotte, V. Nicosia and V. Latora, Phys. Rev. E {\bf 83}, 030103 (2011).

\bibitem{rosvall1} M. Rosvall and C.T. Bergstrom, Proc. Nat. Acad. Sci. (USA) {\bf 104}, 7327 (2007).

\bibitem{rosvall2} M. Rosvall and C.T. Bergstrom, Proc. Nat. Acad. Sci. (USA) {\bf 105}, 1118 (2008).




\bibitem{Ferraro} A. Ferraro, A. Garc\'{i}a-Saez and A. Ac\'{\i}n, Phys. Rev. A {\bf 76}, 052321 (2007).

\bibitem{Garnerone2012b} S. Garnerone, P. Giorda and P.Zanardi, New. J. Phys. {\bf 14} 013011 (2012).

\bibitem{QPercAcin} A. Ac\'{\i}n, J.I. Cirac and M. Lewenstein, Nature Phys. {\bf 3}, 256 (2007).

\bibitem{QPercCuquet1}  M. Cuquet and J. Calsamiglia, Phys. Rev. Lett. {\bf 103}, 240503 (2009).

\bibitem{QPercCuquet2} M. Cuquet and J. Calsamiglia, Phys. Rev. A {\bf 83}, 032319 (2011).

\bibitem{QPercWu}  L. Wu and S.Q. Zhu, Phys. Rev. A {\bf 84}, 052304 (2011).

\bibitem{QSW} S. Perseguers, M. Lewenstein, A. Ac\'{\i}n and J.I. Cirac, Nature Phys. {\bf 6}, 539 (2010). 


\bibitem{muelken} O. Muelken and A. Blumen, Phys. Rep. {\bf 502}, 37 (2011).

\bibitem{Almeida2012} G. M. A. Almeida and A.M.C. Souza, Phys. Rev. A {\bf 87}, 033804 (2013).

\bibitem{QRpaparo} G.D. Paparo and M.A. Mart\'{\i}n-Delgado, Sci. Rep. {\bf 2}, 444 (2012).

\bibitem{QRsilvano} S. Garnerone, P. Zanardi and D.A. Lidar,
  Phys. Rev. Lett. {\bf 108} 230506 (2012).

\bibitem{Garnerone2012} S Garnerone,  Phys. Rev. A {\bf 86}, 032342 (2012)

\bibitem{QRburillo} E. S\'anchez-Burillo, J. Duch, J. G\'omez-Garde\~nes and D. Zueco, Sci. Rep. {\bf 2} 605 (2012).

\bibitem{Caravelli2012} F. Caravelli, A. Hamma, F. Markopoulou and
  A. Riera, Phys. Rev. D {\bf 85}, 044046 (2012)

\bibitem {Sade2005} M. Sade, T. Kalisky, S. Havlin and R. Berkovits,
  Phys. Rev. E {\bf 72} 066123 (2005)

\bibitem{Jahnke2008} L. Jahnke, J. W. Kantelhardt, R. Berkovits and
  S. Havlin Phys. Rev. Lett. {\bf 101} 175702 (2008)

\bibitem{Halu2013} A. Halu, S. Garnerone, A. Vezzani and G. Bianconi,
  Phys. Rev. E {\bf 87} 022104 (2013)

\bibitem{Shapiro1982} B. Shapiro, Phys. Rev. Lett. {\bf 48} 823 (1982) 

\bibitem{spectra} P.  Van Mieghem, {\it Graph Spectra for Complex Networks}, Cambridge University Press, UK (2010).

\bibitem{fn1} {A parameter free Hamiltonian would be 
$H_{\rm network} =\frac{1}{2} \Big (   {\bf p}^{\rm T} {\bf p} + {\bf x}^{\rm T}
2L {\bf  x}\Big )\, $. Such a model contains the, so called,
Goldstone mode as a ground state with a zero momentum state and
independent of the topology, hence the inclusion of an on-site potential. The
coupling parameter $c$ is then needed as comparison between interaction and the on-site
frequency.}.

\bibitem{Agarwal1971} G. Agarwal, Phys. Rev. A {\bf 3}, 828 (1971).


\bibitem{er} P. Erd\H{o}s and A. R\'enyi, Publ. Math. Inst. Hung. Acad. Sci. {\bf 5}, 17 (1960).

\bibitem{conf} M. Molloy and B. Reed, Random Struct. Algorithms {\bf 6}, 161 (1995).

\bibitem{ba} A.L. Barb\'asi and R. Albert, Science {\bf 286}, 509 (1999).

\bibitem{smolin}  T. Konopka, F. Markopoulou and L. Smolin, {\it Quantum Graphity}, arXiv:hep-th/0611197 (2006).

\bibitem{cosmo} D. Krioukov, M. Kitsak, R.S. Sinkovits, D. Rideout, D. Meyer and M. Bogu\~n\'a, Sci. Rep. {\bf 2}, 793 (2012).



   

\end{thebibliography}

\end{document}